\pdfoutput=1
\documentclass[sigconf, nonacm]{acmart}

\AtBeginDocument{%
  }

\usepackage{amsthm}
\usepackage[ruled, vlined, linesnumbered]{algorithm2e}
\usepackage{caption}
\usepackage{subcaption}
\usepackage{multirow}
\usepackage{hyperref}

\usepackage{enumitem}
\setlist{leftmargin=3.5mm}

\newcommand{\IPAs}[0]{\textsc{IPAs}\xspace}
\newcommand{\IPA}[0]{\textsc{IPA}\xspace}
\newcommand{\benchmark}[0]{\textsc{C-Pack-IPAs}\xspace}

\begin{document}

\pagenumbering{arabic}
\setcounter{page}{1}
\pagestyle{plain}
\pagestyle{fancy}
\rhead{P. Orvalho, M. Janota, and V. Manquinho}

\title[\benchmark : A C90 Program Benchmark of \IPAs]{C-Pack of \IPAs: A C90 Program Benchmark of \\Introductory Programming Assignments}

\author{Pedro Orvalho}
\orcid{0000-0002-7407-5967}
\email{pmorvalho@tecnico.ulisboa.pt}
\affiliation{%
  \institution{INESC-ID/IST - U. Lisboa}
  \city{Lisboa}
  \country{Portugal}
}

\author{Mikoláš Janota}
\orcid{0000-0003-3487-784X}
\email{mikolas.janota@cvut.cz}
\affiliation{%
  \institution{Czech Technical University in Prague}
  \city{Prague}
  \country{Czech Republic}
}

\author{Vasco Manquinho}
\orcid{0000-0002-4205-2189}
\email{vasco.manquinho@tecnico.ulisboa.pt}
\affiliation{%
  \institution{INESC-ID/IST - U. Lisboa}
  \city{Lisboa}
  \country{Portugal}
}
\renewcommand{\shortauthors}{P. Orvalho, M. Janota, and V. Manquinho}

\begin{abstract}
Due to the vast number of students enrolled in Massive Open Online Courses (MOOCs), there has been an increasing number of automated program repair techniques focused on introductory programming assignments (\IPAs). Such techniques take advantage of previous correct student implementations in order to provide automated, comprehensive, and personalized feedback to students.

This paper presents \benchmark, a publicly available benchmark of students' programs submitted for 25 different \IPAs. \benchmark contains semantically correct, semantically incorrect, and syntactically incorrect programs plus a test suite for each \IPA. Hence, \benchmark can be used to help evaluate the development of novel semantic, as well as syntactic, automated program repair frameworks focused on providing feedback to novice programmers.
\end{abstract}



\maketitle

\section{Introduction}

Nowadays, thousands of students enroll every year in programming-oriented Massive Open Online Courses (MOOCs)~\cite{clara}. On top of that, due to the current pandemic situation, even small programming courses are being taught online. Providing feedback to novice students in introductory programming assignments (\IPAs) in these courses requires substantial effort and time by the faculty. Hence, there is an increasing need for systems that provide automated, comprehensive, and personalized feedback to students in incorrect programming assignments. Therefore, automated program repair has become crucial to provide automatic personalized feedback to each novice programmer~\cite{drRepair}.

Typically, a programming assignment in Computer Science courses follows a pattern: the lecturer defines a computational problem; students program a solution; each solution is submitted and checked for correctness using pre-defined tests. If the student's tentative solution does not pass a given test, it is deemed incorrect without helpful feedback that would help the student. If a student program does not pass a portion of the pre-defined tests, she usually asks the lecturer for feedback on why her code does not have the expected behavior. If her program does not pass at least one pre-defined test, that means her implementation is semantically incorrect. Unfortunately, it is not feasible to have personalized feedback from a lecturer in many cases due to the growing number of student enrolments. Therefore, automated semantic program repair frameworks~\cite{semFix, directFix, searchRepair, angelix, asr-for-ITSP, clara, verifix, sarfgen,  refactory, refazer} are ideal for providing hints on how students should repair their incorrect programming assignments.

This paper presents \benchmark, a \textbf{C}90 \textbf{P}rogr\textbf{a}m ben\textbf{c}hmar\textbf{k} of introductory programming assignments (\textbf{\IPAs}).
\benchmark is a collection of students' programs submitted for 25 different \IPAs and the test suite used for each \IPA. The set of \IPAs is  described in Section~\ref{sec:IPAs}. For each \IPA, \benchmark has a set of semantically correct and incorrect implementations evaluated with the \ IPA's test suite. \benchmark also contains a set of syntactically faulty programs submitted for each \IPA. 
This paper aims to introduce \benchmark that contains semantically and syntactically incorrect students' implementations. Thus, \benchmark can help evaluate novel semantic, as well as syntactic, automated program repair frameworks whose goal is to assist novice programmers.

\section{\benchmark}
\label{sec:benchmark}

\begin{table}[t!]
\centering
\caption{High-level Description of \benchmark Benchmark.}
\label{tab:table-IPAs}
\scalebox{0.9}{\begin{tabular}{|c|c|c|c|c|}
\hline
\textbf{Labs} &
  \textbf{\#IPAs} &
  \textbf{\begin{tabular}[c]{@{}c@{}}\#Correct \\ Submissions\end{tabular}} &
  \textbf{\begin{tabular}[c]{@{}c@{}}\#Semantically\\ Incorrect \\ Submissions\end{tabular}} &
  \textbf{\begin{tabular}[c]{@{}c@{}}\#Syntactically\\ Incorrect \\ Submissions\end{tabular}}\\ \hline
\textbf{Lab02} & 10          & 316           & 161  & 50 \\ \hline
\textbf{Lab03} & 7           & 145           & 255  & 29 \\ \hline
\textbf{Lab04} & 8           & 192           & 97  & 19 \\ \hline
\textbf{Total} & \textbf{25} & \textbf{653} & \textbf{513} & \textbf{98} \\ \hline
\end{tabular}}
\end{table}

\benchmark is a pack of student programs developed during an introductory programming course in the C programming language. These programs were collected over three distinct practical classes at Instituto Superior Técnico for 25 different \IPAs.
The set of submissions was split into three groups: semantically correct, semantically incorrect, and syntactically incorrect submissions.
The students' submissions that satisfied the set of input-output test cases for each \IPA were considered semantically correct. The submissions that failed at least one input-output test but successfully compiled were considered semantically incorrect implementations. Lastly, the students' submissions that did not successfully compile were considered syntactically incorrect implementations.

\begin{table*}[t!]
\centering
\caption{The number of semantically correct student submissions received for 25 different programming assignments over three lab classes for two different years.}
\label{tab:number-correct-submissions}
\begin{tabular}{ccccccccccccc}
\cline{3-13}
 &
  \multicolumn{1}{l|}{\textbf{}} &
  \multicolumn{1}{c|}{\textbf{E1}} &
  \multicolumn{1}{c|}{\textbf{E2}} &
  \multicolumn{1}{c|}{\textbf{E3}} &
  \multicolumn{1}{c|}{\textbf{E4}} &
  \multicolumn{1}{c|}{\textbf{E5}} &
  \multicolumn{1}{c|}{\textbf{E6}} &
  \multicolumn{1}{c|}{\textbf{E7}} &
  \multicolumn{1}{c|}{\textbf{E8}} &
  \multicolumn{1}{c|}{\textbf{E9}} &
  \multicolumn{1}{c|}{\textbf{E10}} &
  \multicolumn{1}{c|}{\textbf{Total}} \\ \hline
\multicolumn{1}{|c|}{\multirow{3}{*}{\textbf{Year 1}}} &
  \multicolumn{1}{c|}{\textbf{Lab02}} &
  \multicolumn{1}{c|}{25} &
  \multicolumn{1}{c|}{25} &
  \multicolumn{1}{c|}{25} &
  \multicolumn{1}{c|}{23} &
  \multicolumn{1}{c|}{25} &
  \multicolumn{1}{c|}{23} &
  \multicolumn{1}{c|}{22} &
  \multicolumn{1}{c|}{23} &
  \multicolumn{1}{c|}{24} &
  \multicolumn{1}{c|}{23} &
  \multicolumn{1}{c|}{\textbf{238}} \\ \cline{2-13} 
\multicolumn{1}{|c|}{} &
  \multicolumn{1}{c|}{\textbf{Lab03}} &
  \multicolumn{1}{c|}{20} &
  \multicolumn{1}{c|}{17} &
  \multicolumn{1}{c|}{16} &
  \multicolumn{1}{c|}{7} &
  \multicolumn{1}{c|}{16} &
  \multicolumn{1}{c|}{17} &
  \multicolumn{1}{c|}{20} &
  \multicolumn{1}{c|}{-} &
  \multicolumn{1}{c|}{-} &
  \multicolumn{1}{c|}{-} &
  \multicolumn{1}{c|}{\textbf{113}} \\ \cline{2-13} 
\multicolumn{1}{|c|}{} &
  \multicolumn{1}{c|}{\textbf{Lab04}} &
  \multicolumn{1}{c|}{22} &
  \multicolumn{1}{c|}{22} &
  \multicolumn{1}{c|}{19} &
  \multicolumn{1}{c|}{22} &
  \multicolumn{1}{c|}{18} &
  \multicolumn{1}{c|}{19} &
  \multicolumn{1}{c|}{17} &
  \multicolumn{1}{c|}{13} &
  \multicolumn{1}{c|}{-} &
  \multicolumn{1}{c|}{-} &
  \multicolumn{1}{c|}{\textbf{152}} \\ \hline
\multicolumn{1}{l}{} &
  \multicolumn{1}{l}{} &
  \multicolumn{1}{l}{} &
  \multicolumn{1}{l}{} &
  \multicolumn{1}{l}{} &
  \multicolumn{1}{l}{} &
  \multicolumn{1}{l}{} &
  \multicolumn{1}{l}{} &
  \multicolumn{1}{l}{} &
  \multicolumn{1}{l}{} &
  \multicolumn{1}{l}{} &
  \multicolumn{1}{l}{} &
  \multicolumn{1}{l}{} \\ \cline{3-13} 
\textbf{} &
  \multicolumn{1}{l|}{\textbf{}} &
  \multicolumn{1}{c|}{\textbf{E1}} &
  \multicolumn{1}{c|}{\textbf{E2}} &
  \multicolumn{1}{c|}{\textbf{E3}} &
  \multicolumn{1}{c|}{\textbf{E4}} &
  \multicolumn{1}{c|}{\textbf{E5}} &
  \multicolumn{1}{c|}{\textbf{E6}} &
  \multicolumn{1}{c|}{\textbf{E7}} &
  \multicolumn{1}{c|}{\textbf{E8}} &
  \multicolumn{1}{c|}{\textbf{E9}} &
  \multicolumn{1}{c|}{\textbf{E10}} &
  \multicolumn{1}{c|}{\textbf{Total}} \\ \hline
\multicolumn{1}{|c|}{\multirow{3}{*}{\textbf{Year 2}}} &
  \multicolumn{1}{c|}{\textbf{Lab02}} &
  \multicolumn{1}{c|}{13} &
  \multicolumn{1}{c|}{8} &
  \multicolumn{1}{c|}{8} &
  \multicolumn{1}{c|}{7} &
  \multicolumn{1}{c|}{8} &
  \multicolumn{1}{c|}{8} &
  \multicolumn{1}{c|}{7} &
  \multicolumn{1}{c|}{6} &
  \multicolumn{1}{c|}{7} &
  \multicolumn{1}{c|}{6} &
  \multicolumn{1}{c|}{\textbf{78}} \\ \cline{2-13} 
\multicolumn{1}{|c|}{} &
  \multicolumn{1}{c|}{\textbf{Lab03}} &
  \multicolumn{1}{c|}{6} &
  \multicolumn{1}{c|}{5} &
  \multicolumn{1}{c|}{3} &
  \multicolumn{1}{c|}{1} &
  \multicolumn{1}{c|}{4} &
  \multicolumn{1}{c|}{7} &
  \multicolumn{1}{c|}{6} &
  \multicolumn{1}{c|}{-} &
  \multicolumn{1}{c|}{-} &
  \multicolumn{1}{c|}{-} &
  \multicolumn{1}{c|}{\textbf{32}} \\ \cline{2-13} 
\multicolumn{1}{|c|}{} &
  \multicolumn{1}{c|}{\textbf{Lab04}} &
  \multicolumn{1}{c|}{6} &
  \multicolumn{1}{c|}{7} &
  \multicolumn{1}{c|}{6} &
  \multicolumn{1}{c|}{6} &
  \multicolumn{1}{c|}{4} &
  \multicolumn{1}{c|}{4} &
  \multicolumn{1}{c|}{4} &
  \multicolumn{1}{c|}{3} &
  \multicolumn{1}{c|}{-} &
  \multicolumn{1}{c|}{-} &
  \multicolumn{1}{c|}{\textbf{40}} \\ \hline
\end{tabular}
\end{table*}

Table~\ref{tab:table-IPAs} presents the number of submissions gathered. For 25 different programming exercises, this dataset contains
653 different correct programs, 513 semantically incorrect submissions, and 98 syntactically incorrect implementations. 
\benchmark is publicly available at GitHub: \href{https://github.com/pmorvalho/C-Pack-IPAs}{https://github.com/pmorvalho/C-Pack-IPAs}.

Table~\ref{tab:number-correct-submissions} presents the number of correct submissions for each one of the 25 different programming exercises over three lab classes for two different years. In addition, Table~\ref{tab:number-semantically-incorrect-submissions} presents the number of semantically incorrect submissions, and Table~\ref{tab:number-syntactically-incorrect-submissions} shows the distribution of syntactically incorrect submissions for each \IPA.

Furthermore, \benchmark only contains students' submissions that gave their permission to use their programs for academic purposes. Each student's identification was anonymized for privacy reasons, and all the comments were removed from their programs. A unique identifier was assigned to each student. These identifiers are consistent among different \IPAs and different years of the programming course. For example, if the identifier \texttt{stu\_3} appears in more than one programming exercise, it corresponds to the same student. If some students take the course more than once, they are always assigned to the same anonymized identifier.

\begin{table*}[t!]
\centering
\caption{The number of semantically incorrect student submissions received for 25 different programming assignments over three lab classes for two different years.}
\label{tab:number-semantically-incorrect-submissions}
\begin{tabular}{ccccccccccccc}
\cline{3-13}
 &
  \multicolumn{1}{c|}{\textbf{}} &
  \multicolumn{1}{c|}{\textbf{E1}} &
  \multicolumn{1}{c|}{\textbf{E2}} &
  \multicolumn{1}{c|}{\textbf{E3}} &
  \multicolumn{1}{c|}{\textbf{E4}} &
  \multicolumn{1}{c|}{\textbf{E5}} &
  \multicolumn{1}{c|}{\textbf{E6}} &
  \multicolumn{1}{c|}{\textbf{E7}} &
  \multicolumn{1}{c|}{\textbf{E8}} &
  \multicolumn{1}{c|}{\textbf{E9}} &
  \multicolumn{1}{c|}{\textbf{E10}} &
  \multicolumn{1}{c|}{\textbf{Total}} \\ \hline
\multicolumn{1}{|c|}{\multirow{3}{*}{\textbf{Year 1}}} &
  \multicolumn{1}{c|}{\textbf{Lab02}} &
  \multicolumn{1}{c|}{31} &
  \multicolumn{1}{c|}{10} &
  \multicolumn{1}{c|}{7} &
  \multicolumn{1}{c|}{12} &
  \multicolumn{1}{c|}{3} &
  \multicolumn{1}{c|}{5} &
  \multicolumn{1}{c|}{6} &
  \multicolumn{1}{c|}{9} &
  \multicolumn{1}{c|}{21} &
  \multicolumn{1}{c|}{3} &
  \multicolumn{1}{c|}{\textbf{107}} \\ \cline{2-13} 
\multicolumn{1}{|c|}{} &
  \multicolumn{1}{c|}{\textbf{Lab03}} &
  \multicolumn{1}{c|}{32} &
  \multicolumn{1}{c|}{35} &
  \multicolumn{1}{c|}{20} &
  \multicolumn{1}{c|}{67} &
  \multicolumn{1}{c|}{16} &
  \multicolumn{1}{c|}{17} &
  \multicolumn{1}{c|}{8} &
  \multicolumn{1}{c|}{-} &
  \multicolumn{1}{c|}{-} &
  \multicolumn{1}{c|}{-} &
  \multicolumn{1}{c|}{\textbf{195}} \\ \cline{2-13} 
\multicolumn{1}{|c|}{} &
  \multicolumn{1}{c|}{\textbf{Lab04}} &
  \multicolumn{1}{c|}{5} &
  \multicolumn{1}{c|}{11} &
  \multicolumn{1}{c|}{5} &
  \multicolumn{1}{c|}{3} &
  \multicolumn{1}{c|}{10} &
  \multicolumn{1}{c|}{5} &
  \multicolumn{1}{c|}{18} &
  \multicolumn{1}{c|}{10} &
  \multicolumn{1}{c|}{-} &
  \multicolumn{1}{c|}{-} &
  \multicolumn{1}{c|}{\textbf{67}} \\ \hline
 &
   &
  \multicolumn{1}{l}{} &
  \multicolumn{1}{l}{} &
  \multicolumn{1}{l}{} &
  \multicolumn{1}{l}{} &
  \multicolumn{1}{l}{} &
  \multicolumn{1}{l}{} &
  \multicolumn{1}{l}{} &
  \multicolumn{1}{l}{} &
  \multicolumn{1}{l}{} &
  \multicolumn{1}{l}{} &
  \textbf{} \\ \cline{3-13} 
\textbf{} &
  \multicolumn{1}{c|}{\textbf{}} &
  \multicolumn{1}{c|}{\textbf{E1}} &
  \multicolumn{1}{c|}{\textbf{E2}} &
  \multicolumn{1}{c|}{\textbf{E3}} &
  \multicolumn{1}{c|}{\textbf{E4}} &
  \multicolumn{1}{c|}{\textbf{E5}} &
  \multicolumn{1}{c|}{\textbf{E6}} &
  \multicolumn{1}{c|}{\textbf{E7}} &
  \multicolumn{1}{c|}{\textbf{E8}} &
  \multicolumn{1}{c|}{\textbf{E9}} &
  \multicolumn{1}{c|}{\textbf{E10}} &
  \multicolumn{1}{c|}{\textbf{Total}} \\ \hline
\multicolumn{1}{|c|}{\multirow{3}{*}{\textbf{Year 2}}} &
  \multicolumn{1}{c|}{\textbf{Lab02}} &
  \multicolumn{1}{c|}{28} &
  \multicolumn{1}{c|}{2} &
  \multicolumn{1}{c|}{1} &
  \multicolumn{1}{c|}{7} &
  \multicolumn{1}{c|}{2} &
  \multicolumn{1}{c|}{4} &
  \multicolumn{1}{c|}{7} &
  \multicolumn{1}{c|}{2} &
  \multicolumn{1}{c|}{3} &
  \multicolumn{1}{c|}{4} &
  \multicolumn{1}{c|}{\textbf{60}} \\ \cline{2-13} 
\multicolumn{1}{|c|}{} &
  \multicolumn{1}{c|}{\textbf{Lab03}} &
  \multicolumn{1}{c|}{14} &
  \multicolumn{1}{c|}{10} &
  \multicolumn{1}{c|}{11} &
  \multicolumn{1}{c|}{16} &
  \multicolumn{1}{c|}{9} &
  \multicolumn{1}{c|}{6} &
  \multicolumn{1}{c|}{4} &
  \multicolumn{1}{c|}{-} &
  \multicolumn{1}{c|}{-} &
  \multicolumn{1}{c|}{-} &
  \multicolumn{1}{c|}{\textbf{70}} \\ \cline{2-13} 
\multicolumn{1}{|c|}{} &
  \multicolumn{1}{c|}{\textbf{Lab04}} &
  \multicolumn{1}{c|}{6} &
  \multicolumn{1}{c|}{1} &
  \multicolumn{1}{c|}{1} &
  \multicolumn{1}{c|}{2} &
  \multicolumn{1}{c|}{9} &
  \multicolumn{1}{c|}{1} &
  \multicolumn{1}{c|}{4} &
  \multicolumn{1}{c|}{6} &
  \multicolumn{1}{c|}{-} &
  \multicolumn{1}{c|}{-} &
  \multicolumn{1}{c|}{\textbf{30}} \\ \hline
\end{tabular}
\end{table*}

\begin{table*}[t!]
\centering
\caption{The number of syntactically incorrect student submissions received for 25 different programming assignments over three lab classes for two different years.}
\label{tab:number-syntactically-incorrect-submissions}
\begin{tabular}{ccccccccccccc}
\cline{3-13}
 &
  \multicolumn{1}{c|}{\textbf{}} &
  \multicolumn{1}{c|}{\textbf{E1}} &
  \multicolumn{1}{c|}{\textbf{E2}} &
  \multicolumn{1}{c|}{\textbf{E3}} &
  \multicolumn{1}{c|}{\textbf{E4}} &
  \multicolumn{1}{c|}{\textbf{E5}} &
  \multicolumn{1}{c|}{\textbf{E6}} &
  \multicolumn{1}{c|}{\textbf{E7}} &
  \multicolumn{1}{c|}{\textbf{E8}} &
  \multicolumn{1}{c|}{\textbf{E9}} &
  \multicolumn{1}{c|}{\textbf{E10}} &
  \multicolumn{1}{c|}{\textbf{Total}} \\ \hline
\multicolumn{1}{|c|}{\multirow{3}{*}{\textbf{Year 1}}} &
  \multicolumn{1}{c|}{\textbf{Lab02}} &
  \multicolumn{1}{c|}{6} &
  \multicolumn{1}{c|}{0} &
  \multicolumn{1}{c|}{1} &
  \multicolumn{1}{c|}{5} &
  \multicolumn{1}{c|}{4} &
  \multicolumn{1}{c|}{4} &
  \multicolumn{1}{c|}{4} &
  \multicolumn{1}{c|}{2} &
  \multicolumn{1}{c|}{1} &
  \multicolumn{1}{c|}{2} &
  \multicolumn{1}{c|}{\textbf{29}} \\ \cline{2-13} 
\multicolumn{1}{|c|}{} &
  \multicolumn{1}{c|}{\textbf{Lab03}} &
  \multicolumn{1}{c|}{6} &
  \multicolumn{1}{c|}{4} &
  \multicolumn{1}{c|}{1} &
  \multicolumn{1}{c|}{7} &
  \multicolumn{1}{c|}{2} &
  \multicolumn{1}{c|}{1} &
  \multicolumn{1}{c|}{2} &
  \multicolumn{1}{c|}{-} &
  \multicolumn{1}{c|}{-} &
  \multicolumn{1}{c|}{-} &
  \multicolumn{1}{c|}{\textbf{23}} \\ \cline{2-13} 
\multicolumn{1}{|c|}{} &
  \multicolumn{1}{c|}{\textbf{Lab04}} &
  \multicolumn{1}{c|}{2} &
  \multicolumn{1}{c|}{1} &
  \multicolumn{1}{c|}{1} &
  \multicolumn{1}{c|}{0} &
  \multicolumn{1}{c|}{5} &
  \multicolumn{1}{c|}{0} &
  \multicolumn{1}{c|}{1} &
  \multicolumn{1}{c|}{2} &
  \multicolumn{1}{c|}{-} &
  \multicolumn{1}{c|}{-} &
  \multicolumn{1}{c|}{\textbf{12}} \\ \hline
 &
   &
   &
   &
   &
   &
   &
   &
   &
   &
   &
   &
   \\ \cline{3-13} 
\textbf{} &
  \multicolumn{1}{c|}{\textbf{}} &
  \multicolumn{1}{c|}{\textbf{E1}} &
  \multicolumn{1}{c|}{\textbf{E2}} &
  \multicolumn{1}{c|}{\textbf{E3}} &
  \multicolumn{1}{c|}{\textbf{E4}} &
  \multicolumn{1}{c|}{\textbf{E5}} &
  \multicolumn{1}{c|}{\textbf{E6}} &
  \multicolumn{1}{c|}{\textbf{E7}} &
  \multicolumn{1}{c|}{\textbf{E8}} &
  \multicolumn{1}{c|}{\textbf{E9}} &
  \multicolumn{1}{c|}{\textbf{E10}} &
  \multicolumn{1}{c|}{\textbf{Total}} \\ \hline
\multicolumn{1}{|c|}{\multirow{3}{*}{\textbf{Year 2}}} &
  \multicolumn{1}{c|}{\textbf{Lab02}} &
  \multicolumn{1}{c|}{6} &
  \multicolumn{1}{c|}{3} &
  \multicolumn{1}{c|}{0} &
  \multicolumn{1}{c|}{5} &
  \multicolumn{1}{c|}{1} &
  \multicolumn{1}{c|}{6} &
  \multicolumn{1}{c|}{0} &
  \multicolumn{1}{c|}{0} &
  \multicolumn{1}{c|}{0} &
  \multicolumn{1}{c|}{0} &
  \multicolumn{1}{c|}{\textbf{21}} \\ \cline{2-13} 
\multicolumn{1}{|c|}{} &
  \multicolumn{1}{c|}{\textbf{Lab03}} &
  \multicolumn{1}{c|}{1} &
  \multicolumn{1}{c|}{0} &
  \multicolumn{1}{c|}{0} &
  \multicolumn{1}{c|}{1} &
  \multicolumn{1}{c|}{1} &
  \multicolumn{1}{c|}{1} &
  \multicolumn{1}{c|}{2} &
  \multicolumn{1}{c|}{-} &
  \multicolumn{1}{c|}{-} &
  \multicolumn{1}{c|}{-} &
  \multicolumn{1}{c|}{\textbf{6}} \\ \cline{2-13} 
\multicolumn{1}{|c|}{} &
  \multicolumn{1}{c|}{\textbf{Lab04}} &
  \multicolumn{1}{c|}{0} &
  \multicolumn{1}{c|}{0} &
  \multicolumn{1}{c|}{0} &
  \multicolumn{1}{c|}{1} &
  \multicolumn{1}{c|}{1} &
  \multicolumn{1}{c|}{1} &
  \multicolumn{1}{c|}{4} &
  \multicolumn{1}{c|}{0} &
  \multicolumn{1}{c|}{-} &
  \multicolumn{1}{c|}{-} &
  \multicolumn{1}{c|}{\textbf{7}} \\ \hline
\end{tabular}
\end{table*}

\section{\IPAs Description}
\label{sec:IPAs}


The set of \IPAs corresponds to three different lab classes of the introductory programming course to the C programming language at Instituto Superior Técnico. Each lab class focuses on a different topic of the C programming language. 
In Lab02, the students learn how to program with integers, floats, IO operations (mainly \texttt{printf} and \texttt{scanf}), conditionals (if-statements), and simple loops (for and while-loops).
In Lab03, the students learn how to program with loops, nested loops, auxiliary functions, and chars.
Finally, in Lab04, the students learn how to program with integer arrays and strings. 
The textual description of each programming assignment can be found in the public GitHub repository, and the input/output tests used to evaluate semantically the set of students' submissions. Moreover, there is also a reference implementation for each \IPA in the public git repository that can be used by program repair frameworks that only accept a single reference implementation to repair incorrect programs.

\section{Related Work}
\label{sec:related}

Over the last few years, several program repair tools~\cite{clara,sarfgen,refactory, deepfix} 
have exploited diverse correct implementations from previously enrolled students for each \IPA to repair new incorrect student submissions. On the one hand, some syntactic program repair tools~\cite{deepfix, drRepair} have been developed to help students with compilation errors. On the other hand, semantic program repair has also been used to help repair students' programs semantically~\cite{clara,sarfgen,refactory,verifix, autograder}. However, the number of publicly available benchmarks to help develop and evaluate new program repair tools is significantly small. The ITSP dataset~\cite{asr-for-ITSP} has been used by other automated software repair tools~\cite{asr-for-ITSP, verifix-corr} that use only one reference implementation. This dataset is also a collection of C programs although it is well balanced, i.e., the number of correct submissions is closer to the number of incorrect submissions in this dataset. The IntroClass dataset~\cite{introClass-dataset} is a collection of C programs submitted to six different \IPAs and has the information about the number of defects in each program and the total number of unique defects for each \IPA. Codeflaws~\cite{codeFlaws-dataset} is a dataset of programs submitted for programming competitions on the Codeforces website. More program benchmarks are available for other languages than the C programming language. For example, the dataset of Python programs used to evaluate Refactory~\cite{refactory} is also publicly available. More datasets for automated program repair applied to industry software are also available~\footnote{\href{https://program-repair.org/benchmarks.html}{https://program-repair.org/benchmarks.html}}.

\section{Conclusion}
\label{sec:conclusion}
\benchmark, a \textbf{C}90 \textbf{P}rogr\textbf{a}m ben\textbf{c}hmar\textbf{k} of introductory programming assignments (\textbf{\IPAs}), is a publicly available benchmark of students' submissions for 25 different programming assignments. \benchmark has a set of semantically correct and incorrect implementations as well as syntactically faulty programs submitted for each \IPA. 
To the best of our knowledge, \benchmark is one of the few, if not the only, benchmark of \IPAs written in the C programming language that contains both semantically and syntactically incorrect students' implementations and diverse correct implementations for the same \IPA. Thus, \benchmark can help evaluate novel semantic, as well as syntactic, automated program repair frameworks whose goal is to assist novice programmers in introductory programming courses.

\vspace{-4pt}
\begin{acks}
We would like to thank all the students who gave permission to use their programs for our research. Secondly, we would like to thank Instituto Superior Técnico.
This research was supported by Fundação para a Ciência e Tecnologia (FCT) through grant SFRH/BD/07724/2020 and projects UIDB/50021/2020 and PTDC/CCI-COM/32378/2017.
Any opinions, findings, conclusions, or recommendations expressed in this material are those of the author and do not necessarily reflect the views of FCT.
\end{acks}

\bibliographystyle{plain}
\bibliography{mybibliography.bib}

\begin{thebibliography}{10}

\bibitem{verifix-corr}
Umair~Z. Ahmed, Zhiyu Fan, Jooyong Yi, Omar~I. Al{-}Bataineh, and Abhik
  Roychoudhury.
\newblock Verifix: Verified repair of programming assignments.
\newblock {\em CoRR}, abs/2106.16199, 2021.

\bibitem{verifix}
Umair~Z. Ahmed, Zhiyu Fan, Jooyong Yi, Omar~I. Al-Bataineh, and Abhik
  Roychoudhury.
\newblock Verifix: Verified repair of programming assignments.
\newblock {\em ACM Trans. Softw. Eng. Methodol.}, jan 2022.

\bibitem{clara}
Sumit Gulwani, Ivan Radicek, and Florian Zuleger.
\newblock Automated clustering and program repair for introductory programming
  assignments.
\newblock In {\em {PLDI} 2018}, pages 465--480. {ACM}, 2018.

\bibitem{deepfix}
Rahul Gupta, Soham Pal, Aditya Kanade, and Shirish~K. Shevade.
\newblock Deepfix: Fixing common {C} language errors by deep learning.
\newblock In Satinder~P. Singh and Shaul Markovitch, editors, {\em {AAAI}
  2017}, pages 1345--1351. {AAAI} Press, 2017.

\bibitem{refactory}
Yang Hu, Umair~Z. Ahmed, Sergey Mechtaev, Ben Leong, and Abhik Roychoudhury.
\newblock Re-factoring based program repair applied to programming assignments.
\newblock In {\em 34th {IEEE/ACM} International Conference on Automated
  Software Engineering, {ASE} 2019, San Diego, CA, USA, November 11-15, 2019},
  pages 388--398. {IEEE}, 2019.

\bibitem{searchRepair}
Yalin Ke, Kathryn~T. Stolee, Claire~Le Goues, and Yuriy Brun.
\newblock Repairing programs with semantic code search {(T)}.
\newblock In Myra~B. Cohen, Lars Grunske, and Michael Whalen, editors, {\em
  30th {IEEE/ACM} International Conference on Automated Software Engineering,
  {ASE} 2015}, pages 295--306. {IEEE} Computer Society, 2015.

\bibitem{introClass-dataset}
Claire Le~Goues, Neal Holtschulte, Edward~K Smith, Yuriy Brun, Premkumar
  Devanbu, Stephanie Forrest, and Westley Weimer.
\newblock The manybugs and introclass benchmarks for automated repair of c
  programs.
\newblock {\em IEEE Transactions on Software Engineering}, 41(12):1236--1256,
  2015.

\bibitem{autograder}
Xiao Liu, Shuai Wang, Pei Wang, and Dinghao Wu.
\newblock Automatic grading of programming assignments: an approach based on
  formal semantics.
\newblock In Sarah Beecham and Daniela~E. Damian, editors, {\em Proceedings of
  the 41st International Conference on Software Engineering: Software
  Engineering Education and Training, {ICSE} {(SEET)} 2019}, pages 126--137.
  {IEEE} / {ACM}, 2019.

\bibitem{directFix}
Sergey Mechtaev, Jooyong Yi, and Abhik Roychoudhury.
\newblock Directfix: Looking for simple program repairs.
\newblock In Antonia Bertolino, Gerardo Canfora, and Sebastian~G. Elbaum,
  editors, {\em 37th {IEEE/ACM} International Conference on Software
  Engineering, {ICSE} 2015}, pages 448--458. {IEEE} Computer Society, 2015.

\bibitem{angelix}
Sergey Mechtaev, Jooyong Yi, and Abhik Roychoudhury.
\newblock Angelix: scalable multiline program patch synthesis via symbolic
  analysis.
\newblock In Laura~K. Dillon, Willem Visser, and Laurie~A. Williams, editors,
  {\em {ICSE} 2016}, pages 691--701. {ACM}, 2016.

\bibitem{semFix}
Hoang Duong~Thien Nguyen, Dawei Qi, Abhik Roychoudhury, and Satish Chandra.
\newblock Semfix: program repair via semantic analysis.
\newblock In David Notkin, Betty H.~C. Cheng, and Klaus Pohl, editors, {\em
  35th International Conference on Software Engineering, {ICSE} '13}, pages
  772--781. {IEEE} Computer Society, 2013.

\bibitem{refazer}
Reudismam Rolim, Gustavo Soares, Loris D'Antoni, Oleksandr Polozov, Sumit
  Gulwani, Rohit Gheyi, Ryo Suzuki, and Bj{\"{o}}rn Hartmann.
\newblock Learning syntactic program transformations from examples.
\newblock In Sebasti{\'{a}}n Uchitel, Alessandro Orso, and Martin~P. Robillard,
  editors, {\em {ICSE} 2017}, pages 404--415. {IEEE} / {ACM}, 2017.

\bibitem{codeFlaws-dataset}
Shin~Hwei Tan, Jooyong Yi, Yulis, Sergey Mechtaev, and Abhik Roychoudhury.
\newblock Codeflaws: a programming competition benchmark for evaluating
  automated program repair tools.
\newblock In Sebasti{\'{a}}n Uchitel, Alessandro Orso, and Martin~P. Robillard,
  editors, {\em Proceedings of the 39th International Conference on Software
  Engineering, {ICSE} 2017, Buenos Aires, Argentina, May 20-28, 2017 -
  Companion Volume}, pages 180--182. {IEEE} Computer Society, 2017.

\bibitem{sarfgen}
Ke~Wang, Rishabh Singh, and Zhendong Su.
\newblock Search, align, and repair: data-driven feedback generation for
  introductory programming exercises.
\newblock In {\em {PLDI} 2018}, pages 481--495. {ACM}, 2018.

\bibitem{drRepair}
Michihiro Yasunaga and Percy Liang.
\newblock Graph-based, self-supervised program repair from diagnostic feedback.
\newblock In {\em {ICML} 2020}, volume 119 of {\em Proceedings of Machine
  Learning Research}, pages 10799--10808. {PMLR}, 2020.

\bibitem{asr-for-ITSP}
Jooyong Yi, Umair~Z. Ahmed, Amey Karkare, Shin~Hwei Tan, and Abhik
  Roychoudhury.
\newblock A feasibility study of using automated program repair for
  introductory programming assignments.
\newblock In Eric Bodden, Wilhelm Sch{\"{a}}fer, Arie van Deursen, and Andrea
  Zisman, editors, {\em {ESEC/FSE} 2017}, pages 740--751. {ACM}, 2017.

\end{thebibliography}

\end{document}